\documentclass[twocolumn,english]{revtex4}
\usepackage[T1]{fontenc}
\usepackage[utf8]{inputenc}
\usepackage{amsmath}
\usepackage{amssymb}
\usepackage{graphicx}

\makeatletter

\providecommand{\tabularnewline}{\\}

\@ifundefined{textcolor}{}
{%
 \definecolor{BLACK}{gray}{0}
 \definecolor{WHITE}{gray}{1}
 \definecolor{RED}{rgb}{1,0,0}
 \definecolor{GREEN}{rgb}{0,1,0}
 \definecolor{BLUE}{rgb}{0,0,1}
 \definecolor{CYAN}{cmyk}{1,0,0,0}
 \definecolor{MAGENTA}{cmyk}{0,1,0,0}
 \definecolor{YELLOW}{cmyk}{0,0,1,0}
}

\usepackage{babel}

\usepackage{babel}

\makeatother

\usepackage{babel}
\begin{document}

\title{Magnetic dimers and trimers in the disordered $S=3/2$ spin system
BaTi$_{1/2}$Mn$_{1/2}$O$_{3}$}

\author{F. A. Garcia, $^{1}$ U. F. Kaneko, $^{2}$ E. Granado, $^{2}$ J.
Sichelschmidt, $^{3}$ M. H\"olzel, $^{4}$ J. G. S. Duque, $^{5}$
C. A. J. Nunes,$^{6}$ R. P. Amaral, $^{6}$ P. Marques-Ferreira,$^{6}$
R. Lora-Serrano$^{6}$}

\affiliation{$^{1}$IFUSP, Univ. de S\~ao Paulo, 05508-090, S\~ao Paulo-SP, Brasil. }

\affiliation{$^{2}$Inst Fis Gleb Wataghin, Univ. Estadual de Campinas, 13083-970,
Campinas-SP, Brasil. }

\affiliation{$^{3}$Max Planck Institute for Chemical Physics of Solids, D-01187
Dresden, Germany. }

\affiliation{$^{4}$Forschungsneutronenquelle Heinz Maier-Leibnitz (FRM II), Technische
Universit\"at  M\"unchen, Lichtenbergstr. 1, D-85747 Garching, Germany}

\affiliation{$^{5}$Nucleo de F\'isica, Campus Itabaiana, UFS, 49500-000, Itabaiana,
SE, Brazil.}

\affiliation{$^{6}$Univ. Fed. de Uberl\^andia, Instituto de F\'isica, 38400-902,
Uberl\^andia-MG, Brasil.}

\begin{abstract}
We report a structural/magnetic investigation by X-ray absorption
spectroscopy (XAS), neutron diffraction, dc-susceptibility ($\chi_{\mbox{dc}}$)
and electron spin resonance (ESR) of the 12R-type perovskite BaTi$_{1/2}$Mn$_{1/2}$O$_{3}$.
Our structural analysis by neutron diffraction supports the existence
of structural trimers with chemically disordered occupancy of Mn$^{4+}$
and Ti$^{4+}$ ions, with the valence of the Mn ions confirmed by
the XAS measurements. The magnetic properties are explored by combining
dc-susceptibility and $X$-band ($9.4$ GHz) electron spin resonance,
both in the temperature interval of $2\leq T\leq1000$ K. A scenario
is presented under which the magnetism is explained by considering
magnetic dimers and trimers, with exchange constants $J_{a}/k_{B}=200(2)$
K and $J_{b}/k_{B}=130(10)$ K, and orphan spins. Thus, BaTi$_{1/2}$Mn$_{1/2}$O$_{3}$
is proposed as a rare case of an intrinsically disordered $S=3/2$
spin gap system with a frustrated ground state.
\end{abstract}
\email{fgarcia@if.usp.br}

\maketitle
\selectlanguage{english}%

\section{Introduction}

Oxides with perovskite structure are among the most explored systems
in condensed matter physics. These materials are suitable for fundamental
studies in the field, as well as for technological applications, mostly
due to the great number of observed ground states and properties such
as: multiferroicity, high-temperature superconductivity, colossal
magnetoresistance and many others \cite{dagotto_complexity_2005}.

The ideal formula unit for a perovskite compound is $AB$O$_{3}$,
where $A$ is a rare-earth or alkaline earth cation and $B$ is a
smaller transition metal cation (like Ti, Mn, Co or Ni). When a second
($B'$) metal atom is added to the above structure, an ordered double
perovskite-type structure $A_{2}BB'O_{6}$ (where $B$, $B'$ $=3\mathrm{d}$,
$4\mathrm{d}$ and/or $5\mathrm{d}$ metals), as well as the disordered
$AB_{1/2}B'_{1/2}O_{3}$ perovskite structure, might be formed. The
magnetic properties of both spin and orbital degrees of freedom in
these materials are a subject of intense investigation. 

In the case of the $B$-site ordered double perovskites, both the
$B$ and $B'$ sites are arranged in interpenetrating face-centered-cubic
sub lattices. If the $B$ sites are occupied by non magnetic ions
and the $B'$ sites by magnetic ions constrained to interact antiferromagnetically,
the magnetic interactions between the ions at the $B'$ sites will
be geometrically frustrated \cite{aharen_magnetic_2009}. In this
direction, these systems are good platforms for a systematic investigation
of frustrated magnetic interactions, which can give rise to the realization
of exotic ground states such as the spin glasses, spin liquids and
spin ice phases \cite{moessner_geometrical_2006,balents_spin_2010}.
In the past few years, such investigation was carried out for some
of these systems \cite{aharen_magnetic_2009,aharen_magnetic_2010,de_vries_valence_2010,granado_two-dimensional_2013}.

In the case of the disordered structures, geometric frustration is
not certain. In this context, the partially disordered 12R-type perovskite
 BaTi1/2Mn1/2O3 \cite{keith_synthesis_2004} is noteworthy,
 mainly due to its low symmetry and high degree of chemical disorder 
(site occupancy) at the B and B' sites. Indeed, a large Curie-Weiss ($\theta_{\mathrm{CW}}$) constant,
with no sign of a magnetic phase transition down to $T=2$ K, was
reported for this material \cite{keith_synthesis_2004}. Thus, in
BaTi$_{1/2}$Mn$_{1/2}$O$_{3}$ long range magnetic order, if it
occurs, takes place in a temperature $T\ll\theta_{\mathrm{CW}}$,
the hallmark of a frustrated system.

On the other hand, this behavior is not exclusive of geometrically
frustrated systems. In some compounds, due to strong short range interactions,
the spins are coupled and form magnetically correlated states, presenting
an excitation gap between the $S=0$ singlet ground state and the
$S=1$ excited state. These are called spin-gap systems, and the field
has attracted attention since the pioneering work of Nikuni \emph{et
al.}, on the Bosen-Einstein condensation of magnons in the spin-gap
system TlCuCl$_{3}$ \cite{nikuni_bose-einstein_2000}. 

There is a number of $S=1/2$ or $S=1$ spin-gap systems with a $3$D
structure (for instance Ba$_{3}$Cr$_{2}$O$_{8}$ and Ba$_{3}$Mn$_{2}$O$_{8}$
\cite{nakajima_singlet_2006,tsujii_specific_2005}) that have been
proposed and investigated in the past years. However, in BaTi$_{1/2}$Mn$_{1/2}$O$_{3}$,
the Mn cations are all in a $+4$ valence state resulting in a set
of $S=3/2$ spins. Few examples of spin gap systems have been exhibited
in the case of $S=3/2$, among which $R$CrGeO$_{5}$ ($R=$ Y or
$^{154}$Sm) have been recently confirmed as such by an investigation
using inelastic neutron scattering \cite{hase_experimental_2014}.
Furthermore, in BaTi$_{1/2}$Mn$_{1/2}$O$_{3}$ the occupation of
one of the transition-metal sites is suggested to be disordered \cite{keith_synthesis_2004},
making the system also attractive due to the possibility of investigating
the role of disorder in the physics of spin-gap systems \cite{andrade_defect-induced_2012,lavarelo_magnetic_2013}.
In addition, the structure allows the formation of magnetic trimers
which, for an antiferromagnetic exchange, will present a degenarate
ground state, connecting the physics of spin gap and frustrated systems. 

In this work, we report X-rays absorption spectroscopy (XAS), neutron
diffraction, dc-susceptibility and electron spin resonance of the
partially disordered 12R-type perovskite BaTi$_{1/2}$Mn$_{1/2}$O$_{3}$.
Our structural analysis supports a scenario where the magnetism of
the system depends on the occupation of the transition metal sites
at the structural trimers. We propose that at some of these structural
trimers, Mn$^{4+}$ spin dimers and trimers are formed, while a population
of ``orphan'' spins is left behind. Our results strongly suggest
that BaTi$_{1/2}$Mn$_{1/2}$O$_{3}$ is a rare case of a $S=3/2$
spin-gap system, with a  frustrated ground state. In addition, this
is an intrinsically disordered compound.

\section{Experimental}

Polycrystalline samples of BaTi$_{1/2}$Mn$_{1/2}$O$_{3}$ were synthesized
by solid state reaction. Stoichiometric amounts of BaCO$_{3}$, MnO$_{2}$
and TiO$_{2}$ were mixed and grounded in an agata mortar and heated
in air at $900$ $^{0}$C for $24$ hours in tubular furnace. The
material was then regrounded and heated in air at $1100$ $^{0}$C
for $24$ hours. After each grounding/heating step, the sample was
checked by x-ray diffraction to observe the amount of possible spurious
phases. High purity phases could then be obtained by repetition of
the grounding and the heat treatment.

The XAS measurements were carried out at XAFS-1 beamline of the Brazilian
Synchrotron Light Laboratory (LNLS), using Si(111) crystals to monochromatize
the incident beam. A Mn foil spectrum was recorded simultaneously (in
the back channel) so that the edge position of the Mn could be calibrated.
The inflection point in the first resolved peak of the Mn foil was
chosen to be the absorption edge of the metallic Mn ($6539$ eV).
In this way, the edge shifts of the Mn K-edges for all systems in
this work could be compared (i.e. the change in valence of the Mn).
The data were normalized using ATHENA \cite{ravel_<i>athena</i>_2005}.
For a qualitative comparison, reference samples of the Mn oxides MnO
(Mn$^{2+}$), Mn$_{2}$O$_{3}$ (Mn$^{3+}$) and BaMnO$_{3}$ (Mn$^{4+}$)
were used.

A detailed structural investigation was carried out using the high
resolution powder diffractometer SPODI (Structure Powder Diffractometer)
at the Heinz Maier-Leibnitz (FRM II) research reactor, employing a
Ge$511$ monochromator with $\lambda=1.54832(2)$ $\mbox{\AA}$. A
closed cycle cryostat with a base temperature of $T=3.6$ K was employed
for the low-$T$ measurements. The structural refinement of the neutron
powder diffraction data for BaTi$_{1/2}$Mn$_{1/2}$O$_{3}$ was performed
by means of the Rietveld method with the GSAS+EXPGUI package \cite{larson_general_2000,toby_<i>expgui</i>_2001}. 

The Magnetic properties of BaTi$_{1/2}$Mn$_{1/2}$O$_{3}$ were investigated
by dc-susceptibility ($\chi_{\mbox{dc}}$) and $X$-band ($9.4$ GHz
) ESR measurements. In the temperature interval $2\mbox{ K}\leq T\leq350\mbox{ K}$,
$\chi_{\mbox{dc}}$ was measured using a Quantum Design
SQUID magnetometer. 
In the high temperature region, $300\mbox{ K}\leq T\leq1000\mbox{ K},$
$\chi_{\mbox{dc}}$ was measured in Quantum Design SQUID
vibrating sample magnetometer VSM. Care was taken to use similar sample masses
in these measurements and to calibrate the position of the sample
in a way that the moment per mol (f.u.) would render the same result
at $T=300$ K using both equipments. The ESR measurements were performed
in a commercial Bruker Elexsys-500 spectrometer. For the interval
$2\mbox{ K}\leq T\leq300\mbox{ K}$, the temperature was controlled
using a conventional He flow cryostat, whereas for the interval $300\mbox{ K}\leq T\leq900\mbox{ K}$,
a N$_{2}$ flow system was employed.

\section{Results and discussion}

\subsection{XAS and Neutron diffraction}

\begin{figure}
\begin{centering}
\includegraphics[scale=0.32]{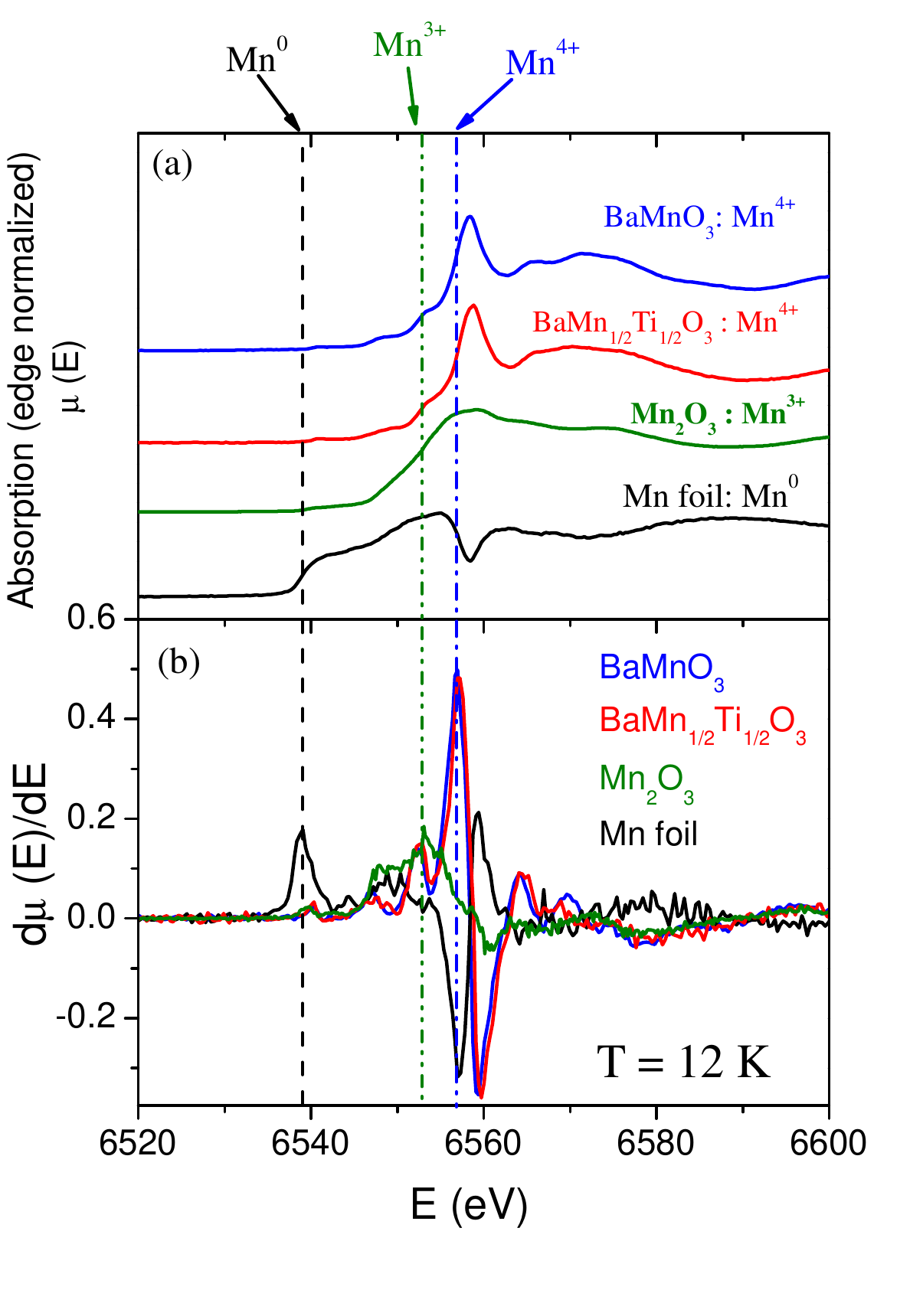} 
\par\end{centering}

\protect\protect\caption{(Color online) (a) X-ray absorption coefficient and (b) first derivatives
of the XAS data (Mn K-edge) measured at $T=12$ K for our BaTi$_{1/2}$Mn$_{1/2}$O$_{3}$
sample (nominal valence of $4+$ ) and the reference samples
(standards) Mn foil, Mn$_{2}$O$_{3}$ (Mn$^{3+}$)
and BaMnO$_{3}$ (Mn$^{4+}$).\label{fig:RLSXAS} }
\end{figure}

The results of our XAS investigation are presented in Fig. \ref{fig:RLSXAS}(a)-(b).
The shapes of the Mn K-edges, as well as the edge shifts ($>2$ eV),
) are clearly distinguishable in Fig. \ref{fig:RLSXAS}(a), indicating
distinct coordination numbers and Mn electronic configurations. Following
the discussion in Refs. \cite{subias_x-ray-absorption_1997,croft_systematic_1997,sikora_x-ray_2006},
the chemical shift observed in the XAS spectra from BaTi$_{1/2}$Mn$_{1/2}$O$_{3}$
and from BaMnO$_{3}$ can be compared, suggesting Mn$^{4+}$ cations
in our BaTi$_{1/2}$Mn$_{1/2}$O$_{3}$ sample. Figure \ref{fig:RLSXAS}(b)
shows  the first energy derivative of the absorption coefficient of
the XAS spectra from the Mn oxides, which reveals the position of the
inflection point of the XAS data in Fig.\ref{fig:RLSXAS}(a). This  gives futher support
for a nearly pure $4+$ valence of the Mn cations in  BaTi$_{1/2}$Mn$_{1/2}$O$_{3}$, 
which, in turn, indicates a negligible level of oxygen vacancies in
this sample.

\begin{figure}
\begin{centering}
\includegraphics[scale=0.28]{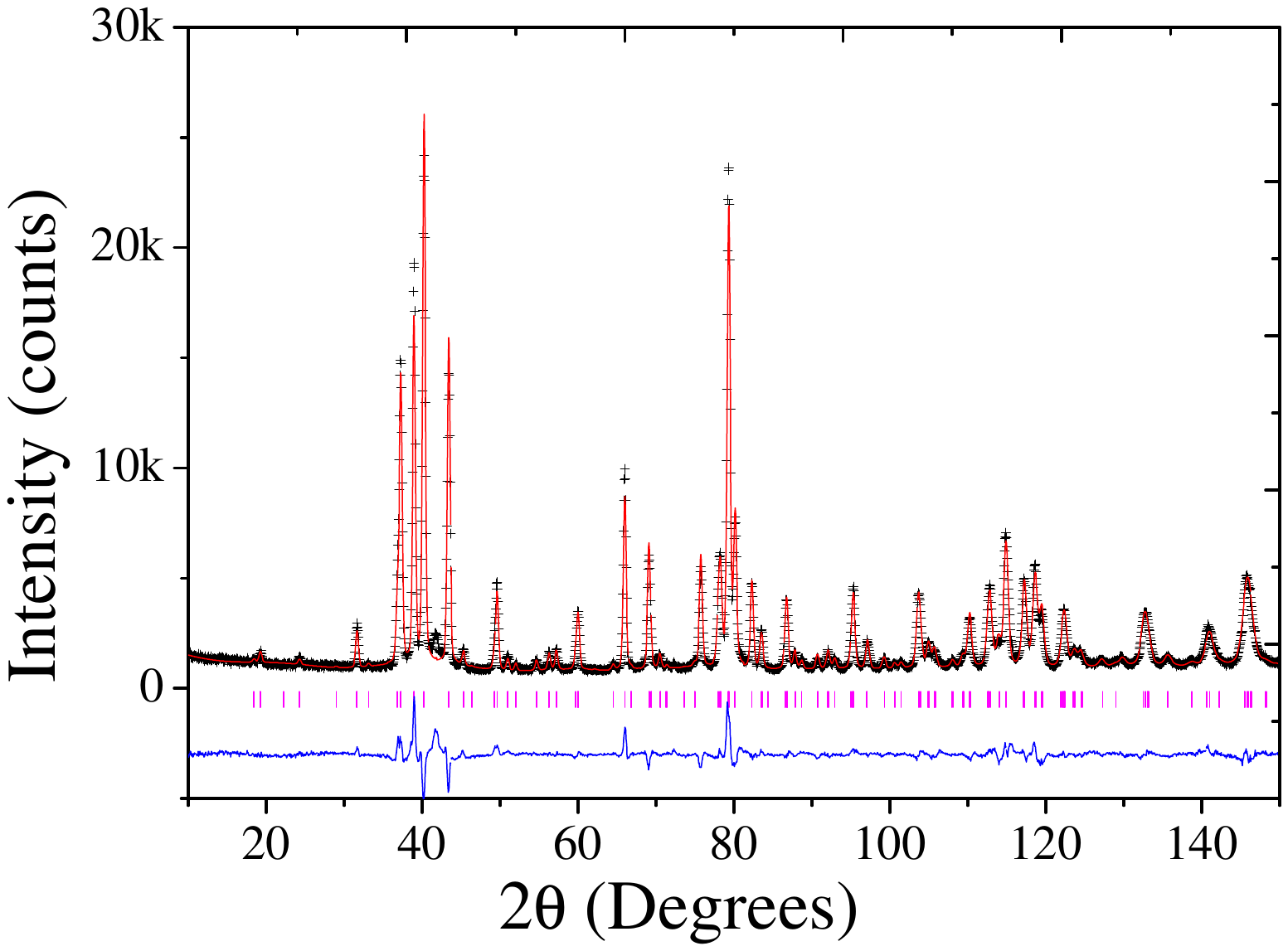} 
\par\end{centering}

\protect\protect\caption{(Color online) High resolution neutron powder diffraction pattern
of BaTi$_{1/2}$Mn$_{1/2}$O$_{3}$. The cross symbols and solid lines
represent observed and calculated patterns, respectively. The difference
curve is shown at the bottom. Vertical bars indicate the expected
Bragg peak positions according to the nuclear structure model.\label{fig:RLSHex} }
\end{figure}

\begin{figure}
\begin{centering}
\includegraphics[scale=0.28]{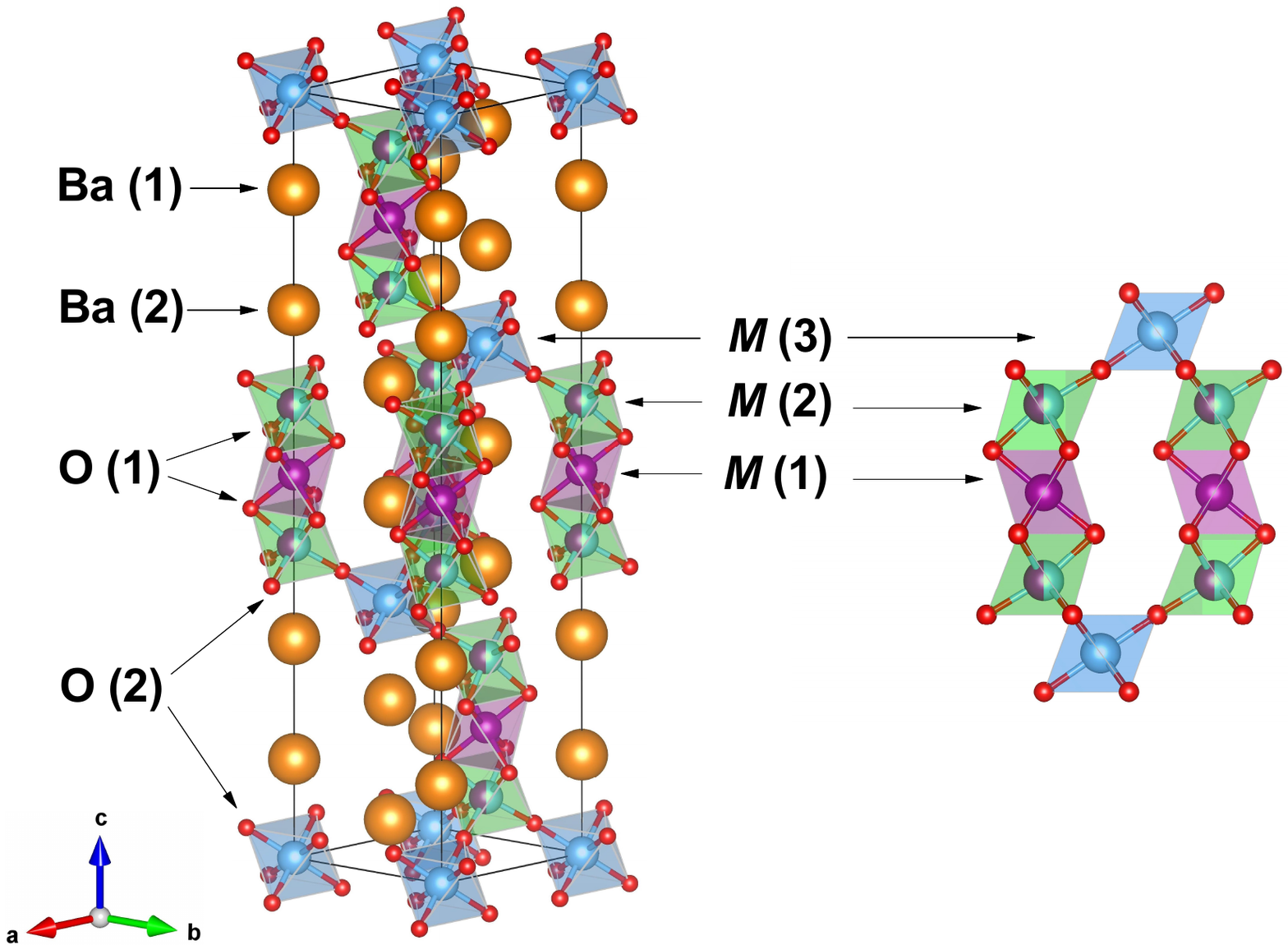} 
\par\end{centering}

\protect\protect\caption{(Color online) Structural model of BaTi$_{1/2}$Mn$_{1/2}$O$_{3}$
with a 12R-type perovskite structure, in which trimers of face-sharing
octahedra lying along the $c$ axis are connected by corner-sharing
octahedra. The face sharing octahedra contains two transition metal
sites $M(1)$ and $M(2)$. The $M(1)$ are occupied only by Mn atoms
(spheres and octahedra in purple), while at the $M(2)$ sites, there
is a mixed occupation of Mn/Ti atoms. The $M(3)$ sites are at the
center of the corner sharing octahedra and are occupied exclusively
by Ti atoms (spheres and octahedra in blue). The O$(1)$ sites label
oxygen sites at the shared face of face-sharing octahedra, whereas
O$(2)$ sites are labeling oxygen sites binding face-sharing to corner-sharing
octahedra. Two sites for barium are indicated as Ba$(1)$ and Ba$(2)$.
\label{fig:RLSstructures} }
\end{figure}

Figure \ref{fig:RLSHex} shows the observed neutron powder diffraction
pattern of BaTi$_{1/2}$Mn$_{1/2}$O$_{3}$ at room temperature (crosses).
A 12R-type perovskite structure was employed in the refinement, see Fig
\ref{fig:RLSstructures}(a). It consists of face-sharing trimers of
octahedra connected by corner-sharing octahedra along the $c$ axis,
under the space group $R\bar{3}m$ \cite{keith_synthesis_2004,miranda_compositionstructureproperty_2009}.
All atomic occupancy factors were fixed at the stoichiometric values
in our Rietveld refinement of the Neutron diffraction data. This procedure
is supported by our XAS investigation and avoids instabilities in
our fit due to correlations of such factors with thermal parameters.

In this structure, there are three independent transition-metal sites
$M$, all of them surrounded by oxygen octahedra (see Fig. \ref{fig:RLSstructures}(a)).
The $M(1)$ and $M(2)$ sites are located at the center and at the
border of the trimers, respectively, while the $M(3)$ site lies outside
the trimers and is in the center of the corner-sharing octahedra connecting
the trimers. The calculated diffraction pattern (line) under this
model is also displayed in Fig. \ref{fig:RLSHex}, showing good agreement
with the experimental data.

The refinement results and relevant bond distances are given in table
\ref{tab:struc-hex}. A good agreement of the atomic positions and
bond distances is found with respect to previously published data
\cite{keith_synthesis_2004,miranda_compositionstructureproperty_2009}.

\begin{table}[ht]
\protect\protect\caption{\label{tab:struc-hex} Refined structural parameters and relevant
bond lengths for BaTi$_{1/2}$Mn$_{1/2}$O$_{3}$ at room temperature.}

\scalebox{0.65}{%
\begin{tabular}{ccccccc}
\hline 
Atom  & Site  & Occupancy  & $x$  & $y$  & $z$  & B$_{iso}$ ($\mbox{\AA}$$^{2}$) \tabularnewline
\hline 
Ba(1)  & 6c  & 1.0  & 0  & 0  & 0.2856(2)  & 0.76(9) \tabularnewline
Ba(2)  & 6c  & 1.0  & 0  & 0  & 0.1290(2)  & 0.12(6) \tabularnewline
M(1)(Mn)  & 3b  & 1.0  & 0  & 0  & 0.5  & 0.0(1) \tabularnewline
M(2)(Ti/Mn)  & 6c  & 0.5/0.5  & 0  & 0  & 0.4091(2)  & 0.2(1) \tabularnewline
M(3)(Ti)  & 3a  & 1.0  & 0  & 0  & 0  & 0.1(1) \tabularnewline
O(1)  & 18f  & 1.0  & 0.1513(2)  & 0.8487(2)  & 0.45656(8)  & 0.42(4) \tabularnewline
O(2)  & 18f  & 1.0  & 0.1673(3)  & 0.8327(3)  & 0.62700(9)  & 0.63(4) \tabularnewline
\hline 
 &  &  & M(1)-O(1) 1.922(2) $\mbox{\AA}$  &  &  & \tabularnewline
 &  &  & M(2)-O(1) 1.995(5) $\mbox{\AA}$  &  &  & \tabularnewline
 &  &  & M(2)-O(2) 1.932(4) $\mbox{\AA}$  &  &  & \tabularnewline
 &  &  & M(3)-O(2) 1.976(3) $\mbox{\AA}$  &  &  & \tabularnewline
\hline 
\end{tabular}} \\

Note: Space group R$\bar{3}$m, $a=5.6910(3)$ $\mbox{\AA}$, $b=5.6910(3)$
$\mbox{\AA}$, $c=27.915(1)$ $\mbox{\AA}$, R$_{wp}$$=8.0\%$, R$_{p}$$=6.0\%$,
$V=783.0(1)$ $\mbox{\AA}{}^{3}$. 
\end{table}

It should be noted that due to the close similarity between the coherent
scattering lengths for Ti and Mn, the relative occupancy of these
atoms in the $M(1)$, $M(2)$ and $M(3)$ transition metal sites (see
Fig. 2) cannot be directly determined from our refinements. However,
this information can be inferred from the bond distances extracted
from the refinements results \cite{keith_synthesis_2004,shannon_revised_1976}.
In fact, in an octahedral coordination, the expected Mn$^{4+}$- O$^{2-}$
and Ti$^{4+}$- O$^{2-}$ covalent bond lengths are $1.930$ and $2.005$
$\mbox{\AA}$ \cite{shannon_revised_1976}, respectively. In Table
\ref{tab:struc-hex} it is shown that the M$(1)$-O$(1)$ bond length
equals $1.922(2)$ $\mbox{\AA}$, being close to the Mn$^{4+}$- O$^{2-}$
bond length, thus suggesting this site is Mn rich. The M$(3)$-O$(2)$
bond length reads $1.976(3)$ $\mbox{\AA}$, being somewhat smaller
than the expected Ti$^{4+}$- O$^{2\textendash}$ bond length.

Therefore, it should be noted that our results do not discard the
possibility of a mixed occupancy at the $M(3)$ site. However, it
is likely that this site is Ti rich. As for the $M(2)$ site, the
$M(2)$-O$(1)$ bond length is $2.005(4)$ $\mbox{\AA}$, while the
$M(2)$-O$(2)$ bond length is $1.925(4)$ $\mbox{\AA}$, consistent
with a mixed occupancy of Mn/Ti atoms at this site. In view of this
analysis, the final refinement was performed assuming a pure occupation
at the $M(1)$ (Mn atoms) and $M(3)$ (Ti atoms) sites, and a mixed
occupation (Mn/Ti atoms) at the $M(2)$ site (see Table \ref{tab:struc-hex}).
This analysis supports the structural model first proposed in Ref.
\cite{keith_synthesis_2004}.

\begin{figure}
\begin{centering}
\includegraphics[scale=0.25]{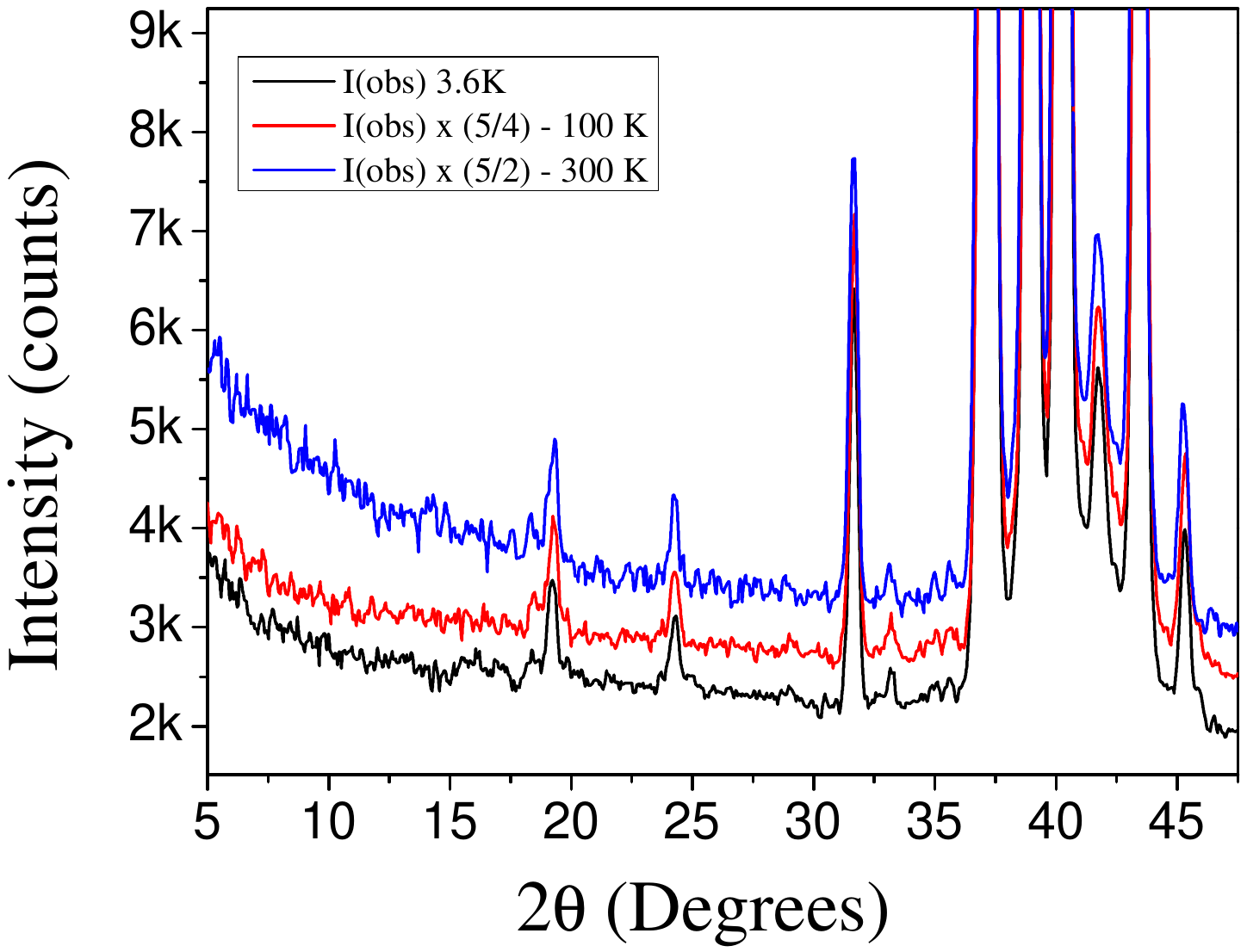} 
\par\end{centering}

\protect\protect\caption{(Color online) Comparison between the observed high resolution neutron
powder diffraction patterns of BaTi$_{1/2}$Mn$_{1/2}$O$_{3}$ measured
at $T=3.6$ K (black line), $T=100$ K (red line) and $T=300$ K (blue
line). Magnifying factors, indicated in the figure, and an off-set
of the vertical axis were used for better visualization. The profile
at $T=300$ K was obtained using a sample changer in air and the upturn
in the very low angle region is due air scattering. No extra features
are observed in the diffraction pattern in this low angle region in
the whole $T$-interval investigated.\label{fig:neutronlowangle} }
\end{figure}

Figure \ref{fig:neutronlowangle} shows in detail the neutron powder
diffraction patterns at low angles for selected temperatures. It can
be seen that no additional feature in the profile is found at low-$T$,
indicating the absence of long range magnetic order down to $T=3.6$
K. This is most likely related to the presence of the Ti$^{4+}$ cations
at the $M(3)$ sites, which breaks the $3$D exchange isolating the
$M(2)-M(1)-M(2)$ trimers from one another. Hence, all the magnetic
properties stem from these isolated, or somewhat weakly coupled, structural
trimers and should be determined by the arrangement of the Mn$^{4+}$
and Ti$^{4+}$ cations at these sites.

\subsection{Susceptibility}

A measurement of $\chi_{\mbox{dc}}$ up to $T=350$ K was previously
reported by Keith \emph{et al} \cite{keith_synthesis_2004}, where
an analysis based on the Curie-Weiss law applied to a restricted $T$-interval
($250\leq T\leq350$) K was presented. Such analysis led to an effective
moment $\mu_{\mbox{eff}}=3.92_{\mbox{B}}$, closely corresponding
to the full moments of Mn$^{4+}$ in the sample.

\begin{figure}
\begin{centering}
\includegraphics[scale=0.27]{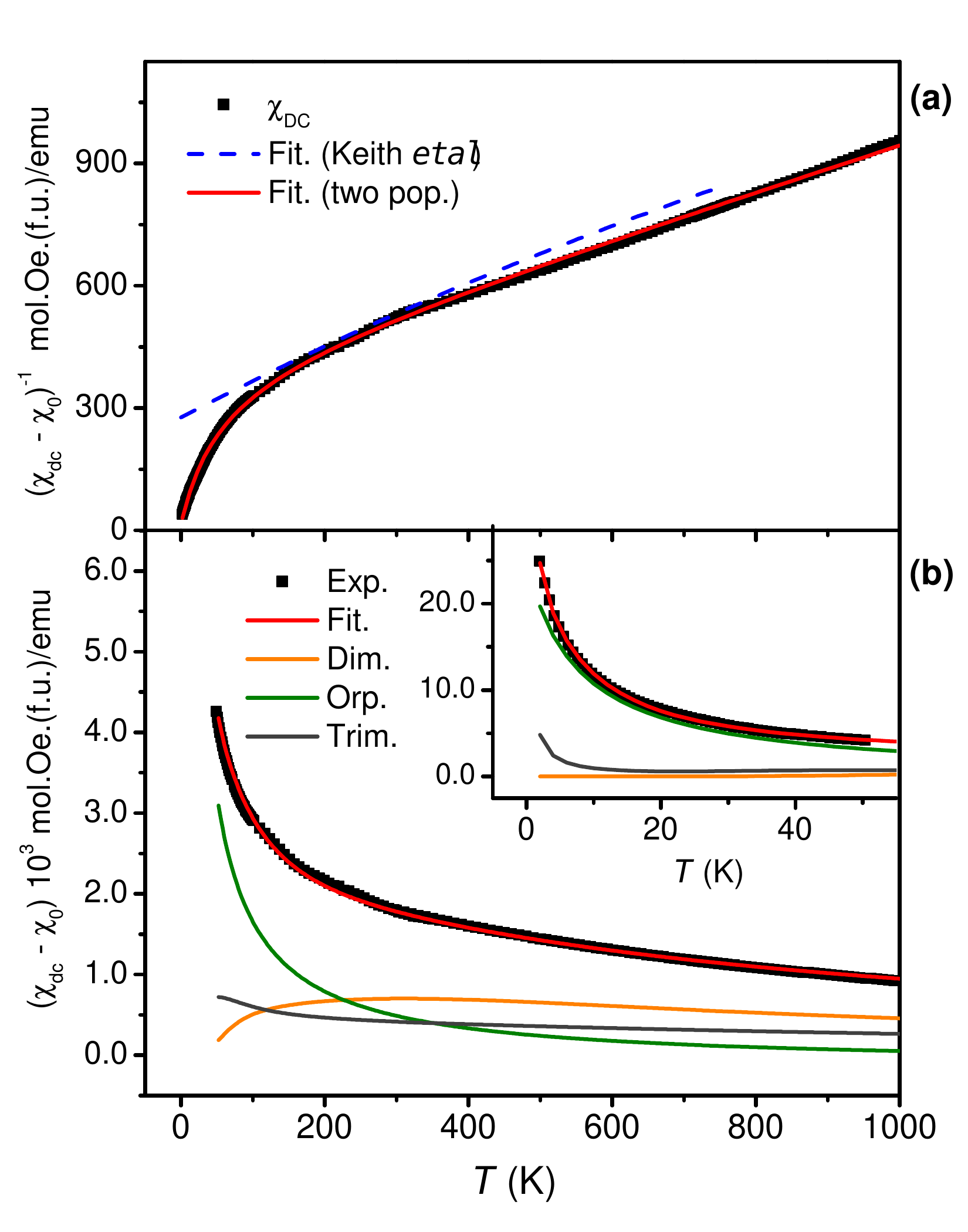} 
\par\end{centering}

\protect\caption{(Color online) Magnetic properties from dc-susceptibility ($\chi_{\mbox{dc}}$)
of BaTi$_{1/2}$Mn$_{1/2}$O$_{3}$. (a) The dashed blue line reproduces
the analysis of Ref. \cite{keith_synthesis_2004}, which describes
the data only in a limited $T-$region; the thick red line is a fitting
of the data to a two population model \cite{schiffer_two-population_1997}.
(b) $(\chi_{\mbox{dc}}-\chi_{0})$ and the fitting of the data to
dimer plus trimer model (thick red line) , along with the separated
contributions $\chi_{\mbox{orp}}$ (thick green line), $\chi_{\mbox{dim }}$
and $\chi_{\mbox{trim}}$ (thick orange and gray lines, respectively).
The inset shows in detail the data and the distinct contributions
for $T\leq50$ K. Results for $\chi_{\mbox{dim }}$  and  $\chi_{\mbox{trim}}$  are obtained
by considering the quantum Hamiltonians  in Eqs. \ref{eq:hamildimer} and \ref{eq:hamiltrimer}. \label{fig:chidcRLS45}}
\end{figure}

 In figure \ref{fig:chidcRLS45}(a), we compare this analysis (dashed-line)
with our experimental results for $(\chi_{\mbox{dc}}-\chi_{0})^{-1}$ , where
$\chi_{0}$ is diamagnetic contribution of the core electrons (see
below), in a wide temperature range.

On top of the deviation observed at the High-$T$ region
, there is a conspicuous drop on $(\chi-\chi_{0})^{-1}$
for $T<200$ K, and the system clearly fails to follow a paramagnetic
behavior in the low-$T$ region. This drop on $(\chi-\chi_{0})^{-1}$
is reminiscent of the overall behavior of the susceptibility of frustrated
systems \cite{moessner_geometrical_2006,balents_spin_2010}. In the
present case, where the magnetism stems from the Mn$^{4+}$ cations
occupying the transition metal sites at the trimers, it is possible
that part of these cations could take part in some correlated spin
state. Hence, one is tempted to adopt a phenomenological two-population
model \cite{schiffer_two-population_1997} to describe our $(\chi-\chi_{0})^{-1}$
data. The data is then fitted (solid line in Fig. \ref{fig:chidcRLS45}(a))
using the following phenomenological model:

\begin{equation}
(\chi_{dc}-\chi_{0})^{-1}=[(\frac{C_{1}}{T+\theta_{1}})+(\frac{C_{2}}{T+\theta_{2}})]^{-1}\label{eq:twopop}
\end{equation}

Where $C_{1}$($C_{2}$) and $\theta_{1}$($\theta_{2}$) stands,
respectively, for the Curie constant and Curie-Weiss constant to be
associated with the correlated (orphan) spins in the system. Since,
ideally, the orphan spins are uncorrelated, one expects $\theta_{2}=0$.
In practice it is usually found that $\theta_{1}\gg\theta_{2}$ \cite{schiffer_two-population_1997}.
The following parameters are obtained: $C_{1}=1.68$ emu.K/mol (f.u.).Oe,
$\theta_{1}=788$ K, $C_{2}=0.12$ emu.K/mol (f.u.).Oe and $\theta_{2}=1.86$
K. The constant contribution $\chi_{0}$ was not fitted. Instead,
we estimated $\chi_{0}=-0.00029$ emu/mol (f.u.) \cite{ashcroft_solid_1976},
and this value was used. Please, observe that we present the data
normalized per formula unit (f.u.) and therefore for one mol of BaTi$_{1/2}$Mn$_{1/2}$O$_{3}$
we have $1/2$ mol of Mn$^{4+}$ cations. 

Since the model offers a fair description of the data, it supports
the existence of orphan and correlated spin populations, to be associated
with distinct energy scales. The nature of the correlated spin state
is rooted in the occupancy of the transition metal sites in the structural
trimers and depends on how all the possible configurations are combined.
The strucutural trimers may be formed by $3$ Mn atoms, $2$ Mn atoms
plus $1$ Ti atom, or $1$ Mn atom plus $2$ Ti atoms. In the first
and second configurations, respectively, we propose that magnetic
trimers and dimers are formed . The interaction takes place by means
of $3$ different exchange paths along the face-sharing octahedra,
allowing a strong magnetic coupling between the $M(1)$ and $M(2)$
sites. Under these assumptions, it can be inferred that the orphan
spins will amount to $1/8$ of Mn$^{4+}$ cations in the sample, while
all other cations will be in some correlated state.

This idea is tested by fitting $\chi_{\mbox{dc}}$ in the entire temperature
range using the following model:

\begin{equation}
\chi_{\mbox{dc}}=\chi_{0}+\chi_{\mbox{orp}}+\chi_{\mbox{dim}}+\chi_{\mbox{trim}}\label{eq:chitotal}
\end{equation}

where $\chi_{\mbox{orp}}$ is the orphan spin contribution and $\chi_{\mbox{dim }}$and
$\chi_{\mbox{trim}}$ are, respectively, the contribution from the
magnetic dimers and trimers. We adopt a Curie-Weiss susceptibility
to model the orphan spin contribution:

\begin{equation}
\chi_{\mbox{orp}}=C/(T+\theta)\label{eq:orphan}
\end{equation}

The contribution of the magnetic dimers and trimers are deduced by
adopting that the interaction among the spins can be described by
Heisemberg-type Hamiltonians. For the dimers we write:

\begin{equation}
\mathcal{H}_{\mbox{dim}}=J_{a}\boldsymbol{S}_{1}.\boldsymbol{S}_{2}\label{eq:hamildimer}
\end{equation}

and, for the trimers:

\begin{equation}
\mathcal{H}_{\mbox{trim}}=J_{a}\boldsymbol{S}_{1}.\boldsymbol{S}_{2}+J_{a}\boldsymbol{S}_{2}.\boldsymbol{S}_{3}+J_{b}\boldsymbol{S}_{1}.\boldsymbol{S}_{3}\label{eq:hamiltrimer}
\end{equation}

In Eqs. \ref{eq:hamildimer} and \ref{eq:hamiltrimer}, $J_{a}$ and
$J_{b}$ are the first and second neighbors exchange constants. In
principle one expects $J_{b}\ll J_{a}$, however, there is large multiplicity
of the exchange paths connecting the transition metal sites (see Fig.
\ref{fig:RLSstructures} ) making possible for $J_{a}$ and $J_{b}$
to be comparable . In this regard, it must be noted that for antiferromagnetic
$J_{a}$ and $J_{b}$, Eq. \ref{eq:hamiltrimer} describes a magnetic
frustrated system. This frustration is important since it reduces
the low temperature magnetic response of the trimer which, otherwise,
would diverge much too fast to be related with our experimental results.
At a qualitative level, the relation between $J_{a}$ and $J_{b}$
can be captured by a classical calculation of the energy of the trimers,
which will depend only on the angles between the spins \cite{white_quantum_2007}:

\begin{figure}
\begin{centering}
\includegraphics[scale=0.55]{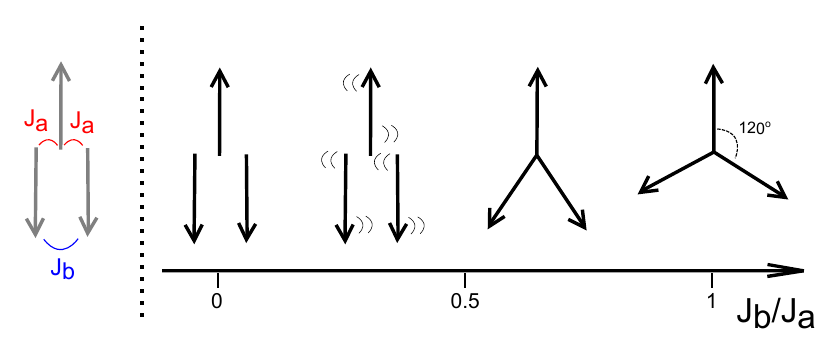} 
\par\end{centering}

\protect\caption{(Color online) Qualitative scenario for the magnetic properties of the trimers.
On the left panel, it is represented that for antiferromagnetic $J_{a}$
and $J_{b}$ the trimer is frustrated. On the right panel, a pictorical
view of the classical calculation is given. It is only for $J_{b}/J_{a}>0.5$
that the spins will not be in a colinear configuration, making the
total spin less than $S=3/2$. The minimum energy and magnetic response
occurs for $J_{b}/J_{a}=1$, and the spins would be found in a triangular
arrangement, for which $S=1/2$. \label{fig:summary}}
\end{figure}

\begin{equation}
E=-2J_{a}(\cos\theta_{12}+\cos\theta_{23}+\frac{J_{b}}{J_{a}}\cos\theta_{13})\label{eq:etrimercla}
\end{equation}

Within this qualitative scenario,  the ground state is obtained by a variational calculation to find
the angles $\theta_{ij}$ which minimizes $E$ as a function of the
fraction $J_{b}/J_{a}$. For $0\leq J_{b}/J_{a}\leq0.5$, one finds
that the spins will be aligned in an antiparallel configuration, with
$J_{b}$ being ineffective at lowering the magnetic response of the
system. However, for $J_{b}/J_{a}>0.5$, the spins will not be colinear.
The lowest possible energy takes place for $J_{b}=J_{a}$, with the
spins aligned in a triangular arrangement. The discussion is summarized
in Fig. \ref{fig:summary}.

Although the classical calculation offers some insight on the problem, the relation between $J_{a}$ and $J_{b}$  in our system must be obtained by considering in full the eigenvalues of the quantum Hamiltonians in Eqs. \ref{eq:hamildimer} and \ref{eq:hamiltrimer}.
The response to the  applied field is calculated by pertubation theory, from which the susceptibilities due to the magnetic dimers and trimers are obtained. The total susceptibility ($\chi_{\mbox{dc}}$, defined in Eq. \ref{eq:chitotal}) is compared  to the data to obtain the values of  $J_{a}$ and $J_{b}$  that best describe the experiment.

 The resulting fitting,
along with the experimental data and the separated contributions from
$\chi_{\mbox{orp}}$ , $\chi_{\mbox{dim }}$and $\chi_{\mbox{trim}}$,
is shown in Fig. \ref{fig:chidcRLS45}(b). The overall picture is
very encouraging. The following parameters are obtained from this
fitting : $\chi_{0}=-0.00014(1)$ emu/mol(f.u.).Oe, $C=0.19(2)$ emu.K/mol(f.u.).Oe,
$\theta=7.7(2)$K, $J_{a}/k_{B}=200(2)$ K and $J_{b}\approx0.65J_{a}$.

The value of $\chi_{0}$ is less than the one estimated theoretically
and may reflect a small, constant, paramagnetic Van Vleck contribution.
The orphan spin Curie constant $C=0.19(4)$ emu.K/mol(f.u.) is close
to the value of $C=0.125$ emu.K/mol(f.u.).Oe, which can be anticipated
for $1/8$ of orphan spins. As expected, $\theta=7.7(2)$ K is a small
energy scale, expressing the correlation among the orphan spins. The
values for the exchange constants,  $J_{a}/k_{B}\approx200$ K and $J_{b}\approx0.65J_{a}$,
are close to values determined for others $3$D spin-gap systems
\cite{deisenhofer_structural_2006,deisenhofer_electron_2012}. The
size of $J_{b}$ is of the correct magnitude to be effective in decreasing
the magnetic response of the trimer. The inset in Fig. \ref{fig:chidcRLS45}(b)
shows in detail the results for $T\leq50$ K.

\subsection{Electron spin resonance}

The single ion physics and qualitative aspects of the effects of
electron correlations can be captured by electron spin resonance
measurements. The ESR parameters of interest are the ESR $g$-values
($g=hv/\mu_{0}H_{\mbox{res}}$), where $H_{\mbox{res}}$ is the resonance
field, and the ESR linewidth, $\Delta H$, which is proportional to
the peak to peak distance of the derivative of the absorption spectrum.

\begin{figure}
\begin{centering}
\includegraphics[scale=0.26]{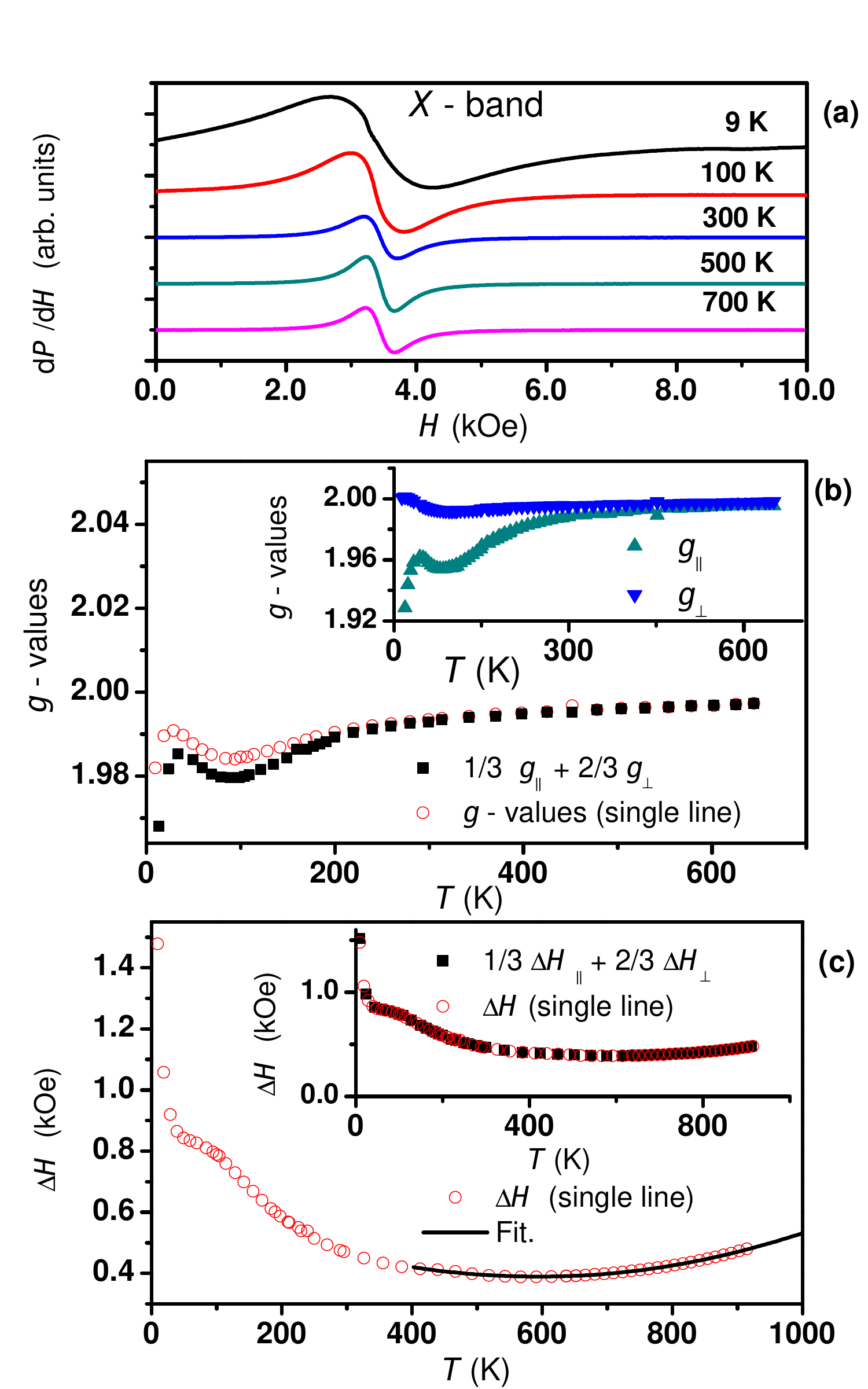} 
\par\end{centering}

\protect\caption{(Color online) $X$-band ($9.4$ GHz) electron spin resonance (ESR)
of BaTi$_{1/2}$Mn$_{1/2}$O$_{3}$. (a) Some representative ESR spectrum
lines. (b) The ESR $g$-values as a function of $T$, obtained from
both a single Lorentzian lineshape analysis and a powder profile fitting.
In the inset, the obtained parameters from the powder profile fitting 
are shown. (c) $\Delta H$, as a function of $T$, obtained from fitting
the spectra by a single Lorentzian lineshape. In the high-$T$ region,
the thick black line corresponds to the fitting of $\Delta H$ by
a model including the usual Kubo-Tomita formula plus an activated
behavior (Eq. \ref{eq:deltaH}). In the inset, we compare the analysis
from the single Lorentzian lineshape and the powder profile fitting.
\label{fig:ESRRLS45}}
\end{figure}

Some representative ESR spectrum lines are shown in Fig. \ref{fig:ESRRLS45}(a).
A careful inspection of the spectra reveals a partial distortion of
some spectrum lines, specially the one for $T=9$ K. This observation
suggests the use of a powder profile fitting since any spin $S>1/2$
would exhibit an anisotropy in a hexagonal lattice. We decided to
use two procedures, fitting the spectra either by a single Lorentzian
lineshape or by a powder profile. A comparison between the two sets
of parameters allows one to determine which aspects of the results
are qualitatively robust. 

By fitting the spectra with a powder profile, we are under the assumption
that the system anisotropy can be described by a $g$-tensor with
two components, one of which is an in-plane $g$-value ($H$ || $ab$,
$g_{||}$) and the other is an out-of-plane $g$-value ($H$ || $c$,
$g_{\perp}$). Experiments on single crystals would allow a proper
investigation of the anisotropy of the ESR parameters which would be
great importance to confirm the existence of dimers and trimers in
the system \cite{deisenhofer_structural_2006}.

In reference to the results presented in Fig. \ref{fig:chidcRLS45}(b),
one expects distinct regimes for the behavior of the ESR parameters.
At the high $T$ region (to be understood as the region at which $T>J_{a}/k_{B}\approx200$
K) the ESR will reflect the local moment characteristics . With lowering
temperature, the dimerizing spins will stop contributing to the
ESR and the response of the magnetic trimer, along with the response
of the orphan spins, will dominate. Hence, one expects a crossover
behavior in the region of intermediate temperatures with energy
scales given by the exchange constants $J_{a}$ and $J_{b}$ \cite{zakharov_spin_2006,deisenhofer_structural_2006,deisenhofer_electron_2012}.

In Fig. \ref{fig:ESRRLS45}(b), we present the ESR $g$-values. In
general, the $g$-values obtained by means of a single Lorentzian
line fitting (full squares) are weakly temperature dependent at high $T$
assuming a value close to $g=1.99$, expected for transition metal
ions with small spin orbit coupling \cite{abragam_electron_1970}.
Around $T=200$ K, and with lowering temperature, one observes
a more conspicuous drop in the $g$-values. These values are compared
with $g'=1/3g_{||}+2/3g_{\perp}$ (open circles), where $g_{||}$
and $g_{\perp}$ are the parameters obtained from the powder fitting.
It is shown that $g\approx1/3g_{||}+2/3g_{\perp}$ in the entire temperature
range. 

In the inset, $g_{||}$ and $g_{\perp}$ are presented separately.
At high $T$ , both parameters converge for $g_{||}\approx g_{\perp}=1.99$.
In the low $T$ region,  $g_{\perp}$ is nearly constant, whereas there
is a clear drop of $g_{||}$.  Therefore, the low-T behavior of the $g$-values is mainly 
due to the T-dependency of  $g_{||}$. Put together, these results suggest that
the system develops an easy axis anisotropy as the temperature decreases,
starting at about $T\approx200$ K. 

This easy-axis anisotropy is most likely related to the formation of the trimers,
 since the energy scale of the correlation between the orphan spins, $\theta=7.7(2)$ K, is significantly
lower. In this direction,  one can see a clear upturn in the $g$-values
at about $T\approx100$ K, which coincides with the temperature at
which the trimer susceptibility (Fig. \ref{fig:chidcRLS45}(b)) become
dominant over the dimer response. Thus, the evolution of the $g$-values can be 
connected to the energy scales given by the exchange constants $J_{a}$ and $J_{b}$.

The evolution of $\Delta H$ with $T$ is presented on Fig. \ref{fig:ESRRLS45}(c).
In concentrated systems of transition metal oxides, a strong exchange
narrowing effect will usually make the spectra from powdered samples
resemble a single Lorentzian lineshape, with some distortion due to
the underlying anisotropy \cite{anderson_exchange_1953,huber_electron_1975}.
In practice, it is usual in these situations to analyze the spin dynamics,
as expressed by $\Delta H$, in terms of a single Lorentzian fitting,
or in terms of the more crude peak-to-peak distance \cite{causa_high-temperature_1998,huber_epr_1999,zakharov_electron_2008}.
Here, we first follow the former phenomenological approach and then
we briefly compare these results with those from a more accurate analysis
using the powder profile fitting . 

At the low $T$ region, one expects that the linewidth due to the
dimers should go to $0$, whereas the response due to orphans and
trimers should diverge. The shoulder in $\Delta H$ around $T=100$
K may be associated with this crossover at which the trimer response
become dominant over the dimer. At high-$T$, the system should follow
the Kubo-Tomita formula, converging for a constant value of $\Delta H_{\infty}^{\mbox{high}-T}$
for $T\gg J_{a}/k_{B}=200$ K. However, a timid increase of $\Delta H$
is observed. We fit this behavior with a phenomenological model that
reads:

\begin{equation}
\Delta H(T)=\frac{(1/T)}{[1/(T+\theta)]}\Delta H_{\infty}^{\mbox{high}-T}+A\exp(-\Delta/k_{B}T)\label{eq:deltaH}
\end{equation}

The first term in Eq. \ref{eq:deltaH} is due to the Kubo-Tomita formula
and expresses the relaxation mechanism due to the exchange coupling
between the spin states in the system. We obtained $\Delta H_{\infty}^{\mbox{high}-T}=282(4)$Oe,
indicating that anisotropic spin interactions are relatively at high $T$, as it is also indicated by the g-values. 
The parameters of the second term are $A=8.9(7)\times10^{3}$
Oe and $\Delta/k_{B}=3832(92)$ K. As it is well known \cite{abragam_electron_1970},
one expects an Orbach mechanism to produce an exponential increase
of the linewidth. However, it is certain that $\Delta/k_{B}=3832(92)$
K is a much too high energy scale to be associated with phonons. This
situation was associated with different Jahn-Teller distortions, which
are close in energy \cite{deisenhofer_structural_2006}. Hence, $\Delta/k_{B}=3832(92)$
K would correspond to the energy barrier separating two Jahn-Teller
distortions. 

The inset of the figure displays a comparison between $\Delta H$,
as obtained from the single line fitting, and the parameters $\Delta H_{||}$
and $\Delta H_{\perp}$, from the powder profile fitting. It is again
suggested that the system has an easy-axis anisotropy (since $\Delta H\approx(1/3)\Delta H_{||}+(2/3)\Delta H_{\perp}$).

We should also comment on some $Q$-band measurements (not shown).
In the presence of an inhomogeneous distribution of Mn sites, one
could expect an inhomogeneous broadening of the resonance. This would
appear as a field dependent broadening of the ESR linewidth. Therefore,
the sample was also measured at $Q$-band to check for these inhomogeneities
and the resonance was found to be homogeneous.

\section{Summary}

The structural and magnetic properties of BaMn$_{1/2}$Ti$_{1/2}$O$_{3}$
were investigated. A scenario was formulated under which the magnetism
of the system stems from the formation of magnetic dimers and trimers
of Mn cations, coexisting with a population of orphan spins. This
scenario resulted in a good description of the susceptibility of the
system in the whole temperature interval. The dimers are formed by
means of a strong first neighbor exchange constant $J_{a}$, whereas
the trimers require a second neighbor exchange constant $J_{b}$ 
which competes with $J_{a}$. This competition introduces frustration
in the magnetic response of the Mn trimers and  was captured by
our ESR experiment, as discussed in the context of the temperature
dependence of the $g$-values. Thus, it is proposed that BaTi$_{1/2}$Mn$_{1/2}$O$_{3}$
is a rare case of an intrinsically disordered $S=3/2$ spin gap system,
with energy scales given by $J_{a}/k_{B}=200(2)$ K and $J_{b}/k_{B}=130(10)$
K. The deduced ground state is magnetic and frustrated, connecting
the physics of spin gap and frustrated systems. Inelastic neutron
scattering experiments would be highly instrumental to test the spin
gap, whereas susceptibility and specific heat at further lower temperatures
are required to investigate the frustrated ground state.

\section{Acknowledgments}

The work is being financed by FAPEMIG (MG-Brazil) Grants No. APQ-01577-09,
APQ-02253-12, 2010-EXA023 and 2012-EXA011, CNPq (Brazil) Grants 308355/2009-1,
482549/2010-6, and PICT 1043. F. A. Garcia would like to acknowledge
FAPESP for the financial support during the work and Eric Andrade
for fruitful discussions on the basics of frustrated systems.


\end{document}